\documentclass[aps,prd,12pt,nofootinbib,floatfix]{revtex4}

\usepackage{amsfonts}
\usepackage{amsmath,epsfig,bm}
\usepackage{graphicx}
\usepackage{color}

\usepackage[setpagesize=false]{hyperref}

\newcommand{\be}{\begin{equation}}
\newcommand{\ee}{\end{equation}}
\newcommand{\ba}{\begin{eqnarray}}
\newcommand{\ea}{\end{eqnarray}}
\newcommand{\bd}{\begin{displaymath}}
\newcommand{\ed}{\end{displaymath}}

\def\thalf{{\textstyle{\frac{1}{2}}}}
\def\tthalf{{\textstyle{\frac{3}{2}}}}
\def\oneth{{\textstyle{\frac{1}{3}}}}
\def\twoth{{\textstyle{\frac{2}{3}}}}

\def\oneqt{{\textstyle{\frac{1}{4}}}}

\def\e{{\mathrm e}}

\newcommand{\mb}[1]{\mathbf{#1}}

\begin{document}

\title{Causal Electric Charge Diffusion and Balance Functions in Relativistic Heavy Ion Collisions}

\author{Joseph I. Kapusta and Christopher Plumberg}
\affiliation{School of Physics \& Astronomy, University of Minnesota, Minneapolis, MN 55455,USA}


\begin{abstract}
We study the propogation and diffusion of electric charge fluctuations in high energy heavy ion collisions using the Cattaneo form for the dissipative part of the electric current.  As opposed to the ordinary diffusion equation this form limits the speed at which charge can propagate.  Including the noise term in the current, which arises uniquely from the fluctuation-dissipation theorem, we calculate the balance functions for charged hadrons in a simple 1+1 dimensional Bjorken hydrodynamical model.  Limiting the speed of propogation of charge fluctuations increases the height and reduces the width of these balance functions when plotted versus rapidity.  We also estimate the numerical value of the associated diffusion time constant from AdS/CFT theory.
\end{abstract}

\maketitle

\section{Introduction}
\label{sec:introduction}

The main motivation for colliding large nuclei at high energies is to produce matter with temperatures greater than 100 MeV that would have existed during the first microsecond after the big bang.  There are large and extensive experimental programs at the Relativistic Heavy Ion Collider (RHIC) at Brookhaven National Laboratory on Long Island, New York and at the Large Hadron Collider (LHC) at CERN in Geneva, Switzerland.  Thousands of hadrons are produced in these collisions which makes the application of statistical mechanics and hydrodynamics plausible, and in fact the standard model of these collisions incorporates these both in principle and in practice.  Of course one should expect significant fluctuations in the observables and these can be put to good use to extract new physics.  For example, critical points are characterized by large fluctuations.  This led to the suggestion to study fluctuations in conserved quantities, such as electric charge, baryon number, and strangeness on an event-by-event basis  \cite{Shuryak1998,Hatta2003}.  Overviews, summaries, and recent progress can be followed via the proceedings of the series of the so-called Quark Matter conferences, the most recent being \cite{QM2014,QM2015,QM2017}. 

In this paper we focus on fluctuations and correlations of the electric charge.  Electric charge is much easier to measure in high energy heavy ion collisions than baryon number or strangeness.  We assume that the net electric charge in the central rapidity region is zero, which is a very good approximation for the top RHIC energies and at the LHC.  Since gradients in temperature, flow velocity, etc. are large it has been found necessary to go to second order viscous fluid dynamics.  Under certain conditions first order viscous fluid dynamics can lead to super-luminal signal propagation, a theoretically unsatisfactory possibility and in addition can lead to instabilities in numerical simulation of high energy nuclear collisions.  The ordinary diffusion equation has instantaneous signal propagation and therefore should be replaced with an equation which respects relativity.  The simplest extension of the ordinary diffusion equation is usually attributed to Cattaneo \cite{Cattaneo} who actually studied heat conduction.  To our knowledge the first application of his equation to heavy ion collisions was by Abdel Aziz and Gavin \cite{Aziz}.  Neglecting noise, the electric charge current which leads to the scalar Cattaneo equation is \cite{Clint2014}
\be
J^{\mu}_Q = n_Q u^{\mu} + \sigma_Q T \Delta^{\mu} (1+\tau_Q u\cdot \partial)^{-1} \left(\frac{\mu_Q}{T}\right) \ .
\label{CatCurrent}
\ee
Here $n_Q$ is the electric charge density with associated charge chemical potential $\mu_Q$.  The charge conductivity is represented by $\sigma_Q$ and the temperature by $T$.  The gradient orthogonal to the flow velocity is
\be
\Delta^{\mu} = \partial^{\mu} -  u^\mu (u\cdot \partial) \,.
\ee
A new time constant $\tau_Q$ naturally appears (which could depend on $T$).  Note that this is not a simple gradient expansion as the derivative appears in the denominator.  The speed of propogation of signals is given by $v_Q^2 = D_Q/\tau_Q$ where $D_Q = \sigma_Q \chi_Q$ is the diffusion constant and $\chi_Q$ is the electric charge susceptibility.  For a given diffusion constant there is a minimum value of $\tau_Q$ for which signals propagate slower than the speed of light.  The ordinary diffusion equation has $\tau_Q \rightarrow \infty$.

The fluctuation-dissipation theorem tells us that along with the dissipation arising from diffusion there are also fluctuations \cite{Landau:1980st}.  Thus on the right side of Eq. (\ref{CatCurrent}) there is added a term $I^{\mu}$ whose average value is zero but which has a nonzero average $\langle I^{\mu}(x_1) I^{\nu}(x_2) \rangle$ which is uniquely determined by the theorem.  The relativistic version was worked out in \cite{Kapusta2012} for the case of ordinary diffusion.  Defining $h^{\mu\nu} = u^{\mu}u^{\nu} - g^{\mu\nu}$ it is
\be
\langle I^{\mu}(x_1) I^{\nu}(x_2) \rangle = 2 \sigma_Q T h^{\mu\nu} \delta (x_1 - x_2) \, .
\ee
This is white noise since the Fourier transform is independent of frequency and wavenumber.  Ordinary electric charge diffusion was applied to the balance functions, which measure two-particle correlations in momentum space \cite{Bass2000,Jeon2002,Pratt2012,Bozek2012}, in \cite{Todd2014}. 

In this paper we extend the study of \cite{Todd2014} to the Cattaneo equation with finite $\tau_Q$.  We will, however, simplify a few of the calculations of Ref. \cite{Todd2014} to focus on the essential physics provided by a finite speed of propagation.  Now the fluctuations are no longer a delta-function in time.  In the local rest frame \cite{Clint2014}
\be
\langle I^i(x_1) I^j(x_2) \rangle = \frac{\sigma_Q T}{\tau_Q} \delta ({\bf x}_1 - {\bf x}_2)  {\rm e}^{-|t_1-t_2|/\tau_Q} \delta_{ij} \, .
\ee 
As $\tau_Q \rightarrow \infty$ this clearly reproduces the white noise of the ordinary diffusion equation.  As in Ref. \cite{Todd2014} we will use 1+1 dimensional boost invariant hydrodynamics  to carry out the calculations as far as possible analytically.  Even then the analysis is more involved because of the memory effects arising from the colored noise.

The outline of the article is as follows.  In Sect. II we work out the relevant equations for charge diffusion and fluctuations in boost invariant hydrodynamics.  In Sect. III we solve the resulting homogeneous equation and in Sect. IV the inhomogeneous equation.  In Sect. V we determine the correlation functions; in particular, we show how to deal with the colored noise in the expanding system.  In order to compare with experimental measurements it is necessary to subtract out self-correlations among the fluid elements.  This is a delicate matter for colored noise and is done in Sect. VI.  Section VII contains numerical results for the correlation functions.  Section VIII contains some straightforward, although not entirely realistic, phenomenological analyses of experimental data.  Section IX compares results from the Cattaneo equation, and the next higher order Gurtin-Pipkin equation \cite{GP,Clint2014}, with results obtained from AdS/CFT to estimate the values of the diffusion constant and relaxation time scales.  Conclusions are provided in Sect. X.

It should be acknowledged that there are other sources of fluctuations in heavy ion collisions, such as initial state fluctuations, fluctuations induced by jets and other high momentum-transfer processes, and fluctuations during hadronization in the final state.  These were surveyed in Ref. \cite{Kapusta2012}.  The ability of hydrodynamic fluctuations to sense a critical point at finite temperature and baryon chemical potential was studied in Ref. \cite{Juan2012} using the ordinary diffusion equation.

\section{Diffusion in Boost Invariant Hydrodynamics}
\label{sec:an-example}

In this section we will derive the relevant equations for charge diffusion and fluctuations during the expansion of hot matter produced in very high energy heavy ion collisions.  For this purpose we will use the 1+1 dimensional boost-invariant (Bjorken) hydrodynamic model, similar to what was done in Ref. \cite{Kapusta2012}.  In addition, we assume zero net charge and neglect the effects of shear and bulk viscosity so that charge diffusion and fluctuations decouple from the shear and bulk modes.  Obviously this model is very simplified.  Nevertheless, it does provide guidance and intuition before one attempts to study the problem with much more sophisticated and numerically intensive 3+1 dimensional viscous fluid dynamics.  

The energy-momentum tensor in ideal fluid dynamics is
\be
T^{\mu\nu}=wu^{\mu}u^{\nu}-Pg^{\mu\nu} \ .
\ee
The shear and bulk viscosities are ignored to focus on the effects of electric conductivity.  In boost-invariant hydrodynamics one expresses the time and location along the beam direction in terms of the proper time $\tau$ and space-time rapidity $\xi$ as
\ba
t &=& \tau \cosh\xi \nonumber \\
z &=& \tau \sinh\xi
\ea
with the inverse relations
\ba
\tau &=& \sqrt{t^2 - z^2} \nonumber \\
\xi &=& \tanh^{-1}(z/t) \, .
\ea
The flow velocity has the nonvanishing components
\ba
u^0 &=& \cosh\xi \nonumber \\
u^3 &=& \sinh\xi \ . 
\ea
The electric charge current which arises from the Cattaneo equation is 
\be
J^{\mu}_Q = n_Q u^{\mu} + \sigma_Q T \Delta^{\mu} (1+\tau_Q u\cdot \partial)^{-1} \left(\frac{\mu_Q}{T}\right) + I^{\mu} \ .
\ee
Here $n_Q$ is the electric charge density with associated charge chemical potential $\mu_Q$.  The charge conductivity is represented by $\sigma_Q$ and the temperature by $T$.  The gradient orthogonal to the flow velocity is
\be
\Delta^{\mu} = \partial^{\mu} - u^\mu (u\cdot \partial) \,.
\ee
The $I^{\mu}$ is a fluctuation, as described in Ref. \cite{Kapusta2012}.  When the electric charge relaxation time constant $\tau_Q$ is zero this expression for the current reduces to the usual one in lowest order viscous fluid dynamics.  Note that when it is not zero the current involves an infinite number of derivatives.  Nevertheless it does represent the Cattaneo equation in a uniform system, which has only first and second order derivatives and which is causal.  It is useful to note that in the Bjorken model used here
\ba
\Delta^0 &=& =-\frac{\sinh\xi}{\tau} \frac{\partial}{\partial \xi} \nonumber \\
\Delta^3 &=& =-\frac{\cosh\xi}{\tau} \frac{\partial}{\partial \xi}
\ea
and
\be
u \cdot \partial = \frac{\partial}{\partial \tau} \,.
\ee
The fluctuating contribution to the current has the structure
\ba
I^0 &=& s(\tau) f(\xi,\tau) \sinh\xi \nonumber \\
I^3 &=& s(\tau) f(\xi,\tau) \cosh\xi
\label{fluctuating_contribution_to_current}
\ea
where $f(\xi,\tau)$ is a random function whose average value is zero.  (The entropy density is factored out so that $f$ is dimensionless.)  Note that $u \cdot J_Q = n_Q$ gives the proper charge density.

The smooth, background fluid equations lead to the simple equations of motion
\be \label{eq:eomfors}
\frac{ds}{d\tau} + \frac{s}{\tau}=0
\ee
and 
\be 
\label{eq:eomforn}
\frac{dn_Q}{d\tau} + \frac{n_Q}{\tau}=0 \ , 
\ee
independent of the specific equation of state.  The solutions are
\be 
\label{eq:sevol}
s(\tau) = s_i \tau_i/\tau
\ee
and 
\be 
\label{eq:nevol}
n_Q(\tau) = n_{Qi} \tau_i/\tau \ ,
\ee
where $s_i$ and $n_{Qi}$ are the entropy and charge densities at some initial time $\tau_i$.  Here we take  $n_{Qi}=0$ so that the average charge density is zero at all subsequent proper times.

Although the average charge density and chemical potential are zero they do fluctuate.  Those fluctuations are related by $\delta n_Q = \chi_Q \delta \mu_Q$ where $\chi_Q$ is the charge susceptibility.  Hereafter we drop the subscript $Q$ on $\delta n_Q$ for notational simplicity.

It is most convenient to use the Fourier transform
\be
\delta n(\xi,\tau) = \int_{\infty}^{\infty} \frac{dk}{2\pi} {\rm e}^{ik\xi} \delta \tilde{n}(k,\tau)
\ee
and similarly for other functions.  Then charge conservation, $\partial \cdot J_Q = 0$, can be expressed most succinctly via the equation
\bd
\frac{\partial^2}{\partial \tau^2} (\tau \delta \tilde{n}) 
+ \left[ \frac{1}{\tau_Q} - \frac{\partial}{\partial \tau} \ln \left(\frac{\chi_Q T D_Q}{\tau}\right) \right] 
\frac{\partial}{\partial \tau} (\tau \delta \tilde{n}) + \frac{v_Q^2 k^2}{\tau^2}  (\tau \delta \tilde{n})
\ed
\be
= -iks \left[ \frac{\partial \tilde{f}}{\partial \tau} + \left( \frac{1}{\tau_Q} - \frac{1}{\tau}
- \frac{\partial}{\partial \tau} \ln \left(\frac{\chi_Q T D_Q}{\tau}\right) \tilde{f} \right)\right] \,.
\label{gencon}
\ee
Here the diffusion constant $D_Q = \sigma_Q/\chi_Q$ has been used.  The speed of propagation of signals in the Cattaneo equation is $v_Q^2 = D_Q/\tau_Q$ \cite{Clint2014}.  The combination $\tau \delta \tilde{n}$ naturally appears because of Eq. (\ref{eq:eomforn}): when the diffusion constant and associated fluctuations in the current are set to zero then $\delta \tilde{n} \sim 1/\tau$ where the constant of proportionality could be anything including zero.  In the limit $\tau_Q \rightarrow 0$ one recovers the usual diffusion equation
\be
\frac{\partial}{\partial \tau} (\tau \delta \tilde{n}) + \frac{D_Q k^2}{\tau^2}  (\tau \delta \tilde{n}) = -iks \tilde{f}
\ee
which was studied in Ref. \cite{Todd2014}.

\section{Solution to the Homogeneous Equation}

To find the solutions to Eq. (\ref{gencon}) we will first find the solutions to the homogeneous equation, which is in the form of a general confluent equation.  We then construct the solutions to the inhomogenous equation by using the method of variation of constants.

According to CFT calcuations, the electric charge diffusion constant should be determined by $D_Q T = 2\pi$ (see Sec. IX).  Lattice calculations \cite{Aarts2015} show indications of that behavior at large $T$; however, in the range $150 < T < 350$ MeV they show that $D_Q T$ is increasing with $T$ in an approximately linear fashion.  Also, there is no information about the temperature dependence of $\tau_Q$.  One should expect from CFT that $\tau_Q T$ would also be constant.  In the absence of further information we will assume $D_Q$ and $\tau_Q$ are both constant within the temperature range given above.  This also implies that $v_Q^2$ is also a constant.  Finally we assume that $\chi_Q T \sim T^3 \sim 1/\tau$. These assumptions are not at all critical to our study of the essential physics, but it does allow for more analytical results and therefore insight and intuition.  Changing to the dimensionless variable $x \equiv \tau/\tau_Q$ the homogenous equation for $\psi(x) \equiv \tau \delta \tilde{n}$ becomes
\be
\ddot{\psi} + \left( 1 + \frac{2}{x} \right) \dot{\psi} + \frac{v_Q^2 k^2}{x^2} \psi = 0 \,.
\label{homoeq}
\ee
The solutions to this equation are
\be
\psi_{\pm} = x^{\lambda_{\pm} - 1/2} {\rm e}^{-x} M\left(\lambda + \tthalf, 2\lambda + 1, x\right)
\ee
with $\lambda_{\pm} = \pm \sqrt{\oneqt - v_Q^2 k^2}$ if $v_Q^2 k^2 < \oneqt$ and $\lambda_{\pm} = \pm i \sqrt{v_Q^2 k^2 - \oneqt}$ if $v_Q^2 k^2 > \oneqt$.  The $M(a,b,x)$ is Kummer's function and satisfies the differential equation
\be
x \frac{d^2M}{dx^2} + (b-x) \frac{dM}{dx} - aM = 0 \,.
\ee
An integral representation is
\be
M(a,b,x) = \frac{\Gamma(b)}{\Gamma(a)\Gamma(b-a)} \int_0^1 dt \, {\rm e}^{xt} t^{a-1} (1-t)^{b-a-1} \,,
\ee
which has the normalization $M(a,b,0) = 1$.  It is related to Whittaker's function $M_{\kappa,\lambda}(x)$, which satisfies the differential equation
\be
\frac{d^2 M_{\kappa,\lambda}}{dx^2} + \left( -\frac{1}{4} + \frac{\kappa}{x} + \frac{\oneqt - \lambda^2}{x^2} \right) M_{\kappa,\lambda} = 0 \,,
\ee
via
\be
M_{\kappa,\lambda}(x) = x^{\thalf + \lambda} {\rm e}^{-x/2} M\left(\lambda - \kappa + \thalf, 2\lambda + 1, x\right) \,.
\ee
When $\tau_Q = 0$ the solution to the homogeneous equation is simply $\psi_H = \exp(D_Q k^2/\tau)$.

\section{Solution to the Inhomogeneous Equation}

The solution to the inhomogeneous equation is written in terms of a Green function as
\be
\tau \delta \tilde{n}(k,\tau) =- \int_{\tau_0}^{\tau} d\tau' s(\tau') \tilde{G}(k;\tau,\tau') \tilde{f}(k,\tau') \,.
\label{gensol}
\ee
Here $\tau_0$ is the starting time of the hydrodynamic expansion.  The entropy density $s(\tau')$ is explicitly factored out for later convenience; the origin of that can be traced to factoring it out from the noise correlator.  To find the Green function we use the method of variation of constants.  It is expressed as
\be
\tilde{G}(k;\tau,\tau') = ik \left[ a_+(k,\tau') \psi_+(k,\tau) + a_-(k,\tau') \psi_-(k,\tau) \right] \,.
\label{G_definition}
\ee
Using this form in Eq. (\ref{gensol}) we substitute it into 
\be
\frac{d^2}{d \tau^2} \left(\tau \delta \tilde{n}\right) + 
\left( \frac{1}{\tau_Q} + \frac{2}{\tau} \right)\frac{d}{d \tau} \left(\tau \delta \tilde{n}\right)
  + \frac{v_Q^2 k^2}{\tau^2} \left(\tau \delta \tilde{n}\right)
= -iks \left[ \frac{d \tilde{f}}{d \tau} + \left( \frac{1}{\tau_Q} + \frac{1}{\tau}\right) \tilde{f} \right] \,.
\ee
This is solved when
\ba
a_+ \dot{\psi}_+ +  a_- \dot{\psi}_- &=& 0 \nonumber \\
a_+ \psi_+ + a_- \psi_- &=& 1 \,.
\label{a_pm_definitions}
\ea
Thus the Green function is
\be
\tilde{G}(k; \tau, \tau') = ik \left[\frac{\psi_+(\tau) \dot{\psi}_-(\tau') - \psi_-(\tau) \dot{\psi}_+(\tau')}
{\psi_+(\tau') \dot{\psi}_-(\tau') - \psi_-(\tau') \dot{\psi}_+(\tau')} \right] \,.
\label{Green}
\ee

When $\tau_Q = 0$ the Green function is easily found to be
\be
\tilde{G}(k; \tau, \tau') = ik \exp\left[ D_Q k^2 \left( \frac{1}{\tau} - \frac{1}{\tau'} \right) \right] \,.
\label{diffusionG}
\ee

\section{Correlation Functions and Noise}

Suppose we are interested in computing the correlator
\ba
\langle  \tau_1 \delta \tilde{n}(k_1,\tau_1)  \, \tau_2 \delta \tilde{n}(k_2,\tau_2) \rangle &=&
\int_{\tau_0}^{\tau_1} d\tau_1' s(\tau_1')
\int_{\tau_0}^{\tau_2} d\tau_2' s(\tau_2') \, 
\tilde{G}(k_1; \tau_1,\tau_1') \tilde{G}(k_2; \tau_2,\tau_2')  \nonumber \\
&\times& \langle \tilde{f}(k_1,\tau_1') \tilde{f}(k_2,\tau_2') \rangle \,.
\ea
The correlator for the fluctuations is generally written as
\be
\langle \tilde{f}(k_1,\tau_1') \tilde{f}(k_2,\tau_2') \rangle = 2\pi {\cal N}(\tau_1',\tau_2') \delta (k_1 + k_2) \,.
\ee
The function ${\cal N}(\tau_1',\tau_2')$ depends on whether one implements noise from the ordinary diffusion equation or from the Cattaneo equation. 
In either case one gets
\ba
\langle  \tau_1 \delta \tilde{n}(k_1,\tau_1)  \, \tau_2 \delta \tilde{n}(k_2,\tau_2) \rangle &=&
2\pi \delta(k_1+k_2) \int_{\tau_0}^{\tau_1} d\tau_1' s(\tau_1')
\int_{\tau_0}^{\tau_2} d\tau_2' s(\tau_2') \, {\cal N}(\tau_1',\tau_2') \nonumber \\
&\times& \tilde{G}(k_1; \tau_1,\tau_1') \tilde{G}(-k_1; \tau_2,\tau_2')
\label{corrk}
\ea
in $k$-space and
\ba
\langle  \tau_1 \delta n(\xi_1,\tau_1)  \, \tau_2 \delta n(\xi_2,\tau_2) \rangle &=&
\int_{\tau_0}^{\tau_1} d\tau_1' s(\tau_1')
\int_{\tau_0}^{\tau_2} d\tau_2' s(\tau_2') \, {\cal N}(\tau_1',\tau_2') \nonumber \\
&\times& \int \frac{dk}{2\pi} {\rm e}^{ik(\xi_1-\xi_2)} \tilde{G}(k; \tau_1,\tau_1') \tilde{G}(-k; \tau_2,\tau_2')
\label{corrxi}
\ea
in $\xi$-space.

\subsection{White Noise} 

For white noise, the current fluctuations have the form \cite{Kapusta2012}
\be
\langle I^\mu(x_1) I^\nu(x_2) \rangle = 2 \sigma T h^{\mu\nu} \delta^4(x_1-x_2)\, ,
\ee
where $h^{\mu\nu} = g^{\mu\nu} - u^\mu u^\nu$.  This means that for the Bjorken hydrodynamics
\be
\langle f(x_1) f(x_2) \rangle = \frac{2 \sigma T}{s^2} \delta^4(x_1-x_2)\, ,
\ee
by virtue of Eq. (\ref{fluctuating_contribution_to_current}).  After converting to Bjorken coordinates and Fourier-transforming in $\xi$-space, this reads
\be
\langle \tilde f(\tau_1, k_1) \tilde f(\tau_2, k_2) \rangle = \frac{4 \pi \sigma(\tau_1) T(\tau_1)}{A \tau_1 s^2(\tau_1)} \delta(\tau_1-\tau_2) \delta(k_1+k_2) \, ,
\label{white_noise_ff_correlator}
\ee
where $A$ is the transverse area.  This leads to
\be
{\cal N}(\tau_1',\tau_2') = \frac{2 \sigma(\tau_1') T(\tau_1')}{A \tau_1' s^2(\tau_1')} \delta(\tau_1'-\tau_2') \,.
\label{calNwhite}
\ee

\subsection{Cattaneo Noise}

The situation for colored Cattaneo noise is more complicated because it is nonlocal in time.  The $f$ correlator satisfies the equation \cite{Clint2013}
\be
\langle (1 + \tau_Q \partial/\partial \tau_1) \tilde f(\tau_1, k_1) (1 + \tau_Q \partial/\partial \tau_2)\tilde f(\tau_2, k_2) \rangle
 = N(\tau_1) \delta(\tau_1-\tau_2) \delta(k_1+k_2)
\ee
where 
\be
N(\tau) = \frac{4 \pi \sigma(\tau) T(\tau)}{A \tau s^2(\tau)} \,.
\ee
In frequency space
\be
\langle \tilde f(\omega_1, k_1) \tilde f(\omega_2, k_2) \rangle = \frac{ \delta(k_1+k_2) \tilde{N}(\omega_1 + \omega_2)}
{(1 + i\tau_Q \omega_1) (1 + i\tau_Q \omega_2)} \,.
\ee
Now suppose that for some observable $X$
\be
\tilde{X}(k,\tau) =- \int_{\tau_0}^{\tau} d\tau' s(\tau') \tilde{G}_X(k;\tau,\tau') \tilde{f}(k,\tau') \,.
\ee
Then the correlation function for observables $X$ and $Y$ would be
\ba
\langle \tilde{X}(k_1,\tau_1) \tilde{Y}(k_2,\tau_2) \rangle &=& \delta(k_1+k_2)
\int_{\tau_0}^{\tau_1} d\tau_1' s(\tau_1') \tilde{G}_X(k_1;\tau_1,\tau_1')
\int_{\tau_0}^{\tau_2} d\tau_2' s(\tau_2') \tilde{G}_Y(k_2;\tau_2,\tau_2') \nonumber \\
&& \times \int_{-\infty}^{\infty} \frac{d\omega_1}{2\pi} {\rm e}^{i\omega_1 \tau_1'} 
\int_{-\infty}^{\infty} \frac{d\omega_2}{2\pi} {\rm e}^{i\omega_2 \tau_2'}
\frac{ \tilde{N}(\omega_1 + \omega_2)}
{(1 + i\tau_Q \omega_1) (1 + i\tau_Q \omega_2)} \,.
\ea

To evaluate the double integral over $\omega_1$ and $\omega_2$ we change variables to $\bar{\omega} = (\omega_1 + \omega_2)/2$ and $\Delta \omega = \omega_2 - \omega_1$.  Then the double integral over the $\omega_i$ is
\be
\frac{1}{\tau_Q^2} {\rm e}^{-|\tau_2' - \tau_1'|/\tau_Q} \int_{\tau_0}^{{\rm min}(\tau_1',\tau_2')} d\tau N(\tau) {\rm e}^{-2[ {\rm min}(\tau_1',\tau_2')-\tau]/\tau_Q} \,.
\label{doublew}
\ee
The integration over $\tau$ can be thought of as running over the history of the system, prior to the earlier of $\tau_1$ and $\tau_2$, beginning at the initial time $\tau_0$.  (It is assumed that $N(\tau)$ vanishes for $\tau < \tau_0$.)  The exponential kernel in this integration is a direct consequence of using Cattaneo-type diffusion, with states of the medium in the more recent past (nearer to $\min(\tau_1, \tau_2)$) being more heavily weighted than states in the more distant past.

In what follows we will specialize to the case where $N(\tau)$ is a constant.  Then
\be
{\cal N}(\tau_1',\tau_2') = \frac{2 \sigma(\tau_f) T_f}{A \tau_f s^2(\tau_f)} \, \frac{1}{2\tau_Q}
\left[ {\rm e}^{-|\tau_1' - \tau_2'|/\tau_Q} - {\rm e}^{-(\tau_1' + \tau_2' - 2\tau_0)/\tau_Q} \right] \,.
\label{calN}
\ee
In the limit $\tau_Q \rightarrow 0$ this obviously reduces to Eq. (\ref{calNwhite}), as it should.

\section{Self-Correlations}

It is interesting to ask what happens with the ordinary diffusion equation where $\tau_Q = 0$.  In that case, using Eq. (\ref{diffusionG}) one can easily perform the integration over $\tau'$ first, leaving the integration over $k$ to last.  The result is
\be
\langle \delta n(\xi_1,\tau_f)  \, \delta n(\xi_2,\tau_f) \rangle = \frac{\chi_Q(\tau_f) T_f}{A \tau_f}
\left[ \delta (\xi_1 - \xi_2) - \frac{1}{\sqrt{\pi w^2}} {\rm e}^{- (\xi_1 - \xi_2)^2/w^2} \right]
\ee
with $w^2 = 8 D_Q (\tau_0^{-1} - \tau_f^{-1})$.  Note that $\delta (\xi_1 - \xi_2)/A \tau$ in Bjorken hydrodynamics is the equivalent of $\delta ({\bf x}_1 - {\bf x}_2)$ for a static system.  In Refs. \cite{Pratt2012,Todd2014} it was argued that this delta function contribution should be subtracted as it corresponds to a correlation between a particle (or fluid cell) with itself.  
An alternative route to extract the $\delta$-function in $\xi_1-\xi_2$ in pure diffusion is to perform the $\tau'$ integration by parts, as was done in Ref. \cite{Todd2014}.  Due to the complexity of the Cattaneo case this is the route that we will use now.

In rapidity space, the density-density correlator is given by Eq. (\ref{corrxi}).  By construction, the Green's function $\tilde G (k; \tau, \tau')$ satisfies the equation
\bd
\frac{\partial^2}{\partial \tau^2} \tilde G (k; \tau, \tau')
+ \left( \frac{1}{\tau_Q} + \frac{2}{\tau} \right)
\frac{\partial}{\partial \tau} \tilde G (k; \tau, \tau') + \frac{v_Q^2 k^2}{\tau^2}  \tilde G (k; \tau, \tau') = 0 \, .
\label{homogeneous_equation}
\ed
It is useful to note that this same Green's function satisfies a similar differential equation in its second proper time index $\tau'$:
\be
\frac{\partial^2}{\partial \tau'^2} \tilde G (k; \tau, \tau')
- \frac{1}{\tau_Q} \frac{\partial}{\partial \tau'} \tilde G (k; \tau, \tau')
+ \frac{v_Q^2 k^2}{\tau'^2}  \tilde G (k; \tau, \tau') = 0 \, ,
\label{alternate_homogeneous_equation}
\ee
as may be readily checked by direct substitution of Eq. (\ref{Green}) into Eq. (\ref{alternate_homogeneous_equation}).  If we substitute Eq. (\ref{Green}) explicitly into Eq. (\ref{corrxi}) we obtain
\begin{eqnarray}
\langle \delta n(\xi_1,\tau_f) \, \delta n(\xi_2,\tau_f) \rangle
&=& \frac{1}{\tau_f^2 }\int_{\tau_0}^{\tau_f} d\tau_1' s(\tau_1')
\int_{\tau_0}^{\tau_f} d\tau_2' s(\tau_2') \, {\cal N}(\tau_1',\tau_2') \nonumber\\
&\times & \int \frac{dk}{2\pi} \e^{ik(\xi_1-\xi_2)} \tilde{G}(k; \tau_f, \tau_1') \tilde{G}(-k; \tau_f, \tau_2') \nonumber\\
&=& \frac{1}{\tau_f^2 } \int_{\tau_0}^{\tau_f} d\tau_1' s(\tau_1')
\int_{\tau_0}^{\tau_f} d\tau_2' s(\tau_2') \, {\cal N}(\tau_1',\tau_2') \e^{ik(\xi_1-\xi_2)} \nonumber\\
& \times & k^2 \left[ \frac{\psi_+(k, \tau_f) \dot\psi_-(k, \tau_1')-\psi_-(k, \tau_f) \dot\psi_+(k, \tau_1')}{\psi_+(k, \tau_1') \dot\psi_-(k, \tau_1')-\psi_-(k, \tau_1') \dot\psi_+(k, \tau_1')} \right] \nonumber\\
& \times & \left[ \frac{\psi_+(-k, \tau_f) \dot\psi_-(-k, \tau_2')-\psi_-(-k, \tau_f) \dot\psi_+(-k, \tau_2')}{\psi_+(-k, \tau_2') \dot\psi_-(-k, \tau_2')-\psi_-(-k, \tau_2') \dot\psi_+(-k, \tau_2')} \right] \,.
\label{corr_xi_expanded}
\end{eqnarray}

We can eliminate the factor of $k^2$ by making use of Eq. (\ref{alternate_homogeneous_equation}).
On account of Eq. (\ref{a_pm_definitions})
\be
\dot a_+(k, \tau) \psi_+(k, \tau) + \dot a_-(k, \tau) \psi_-(k, \tau) = 0 \,.
\label{useful_properties_of_a_pm}
\ee
This fact allows us to write Eq. (\ref{corr_xi_expanded}) in the somewhat simpler form
\begin{eqnarray}
\langle \delta n(\xi_1,\tau_f) \, \delta n(\xi_2,\tau_f) \rangle
&=& \frac{1}{\tau_f^2} \int_{\tau_0}^{\tau_f} d\tau_1' s(\tau_1')
\int_{\tau_0}^{\tau_f} d\tau_2' s(\tau_2') \, {\cal N}(\tau_1',\tau_2') \int \frac{dk}{2\pi} \e^{ik(\xi_1-\xi_2)} \nonumber\\
& \times & \frac{\tau_1'^2}{v_Q^2} \left( \frac{1}{\tau_Q} \frac{\partial}{\partial \tau_1'} - \frac{\partial^2}{\partial \tau_1'^2} \right)
\left[ \psi_+(k, \tau_f) a_+(k, \tau_1') + \psi_-(k, \tau_f) a_-(k, \tau_1') \right] \nonumber\\
& \times & \left[ \psi_+(-k, \tau_f) a_+(-k, \tau_2') + \psi_-(-k, \tau_f) a_-(-k, \tau_2') \right] \, .
\label{simplified_correlator}
\end{eqnarray}
Focusing on the integral over $\tau_1'$, we integrate by parts.  The term involving the first derivative of $\tau_1'$ becomes
\begin{eqnarray}
&& \frac{1}{\tau_Q} \int_{\tau_0}^{\tau_f} d\tau_1' s(\tau_1')\, {\cal N}(\tau_1',\tau_2') \tau_1'^2
\frac{\partial}{\partial \tau_1'} \left[ \psi_+(k, \tau_f) a_+(k, \tau_1') + \psi_-(k, \tau_f) a_-(k, \tau_1') \right] \nonumber\\
&=& \frac{\tau_f^2}{\tau_Q} s(\tau_f)\, {\cal N}(\tau_f,\tau_2') 
- \frac{\tau_0^2}{\tau_Q} s(\tau_0)\, {\cal N}(\tau_0,\tau_2') \left[ \psi_+(k, \tau_f) a_+(k, \tau_0) + \psi_-(k, \tau_f) a_-(k, \tau_0) \right] \nonumber\\
&-& \frac{1}{\tau_Q} \int_{\tau_0}^{\tau_f} d\tau_1'
\frac{\partial}{\partial \tau_1'}  \left[ s(\tau_1')\, {\cal N}(\tau_1',\tau_2') \tau_1'^2 \right]
\left[ \psi_+(k, \tau_f) a_+(k, \tau_1') + \psi_-(k, \tau_f) a_-(k, \tau_1') \right] \,.
\end{eqnarray}
Notice that the term evaluated at $\tau_1' = \tau_f$, is independent of $k$ due to Eq. (\ref{a_pm_definitions}).  This is therefore a term contributing to self-correlations.

Similar manipulations apply to the term containing the second derivative in Eq. (\ref{simplified_correlator}).  For this term integration by parts yields
\begin{eqnarray}
&& -\int_{\tau_0}^{\tau_f} d\tau_1' s(\tau_1')\, {\cal N}(\tau_1',\tau_2') \tau_1'^2
\frac{\partial^2}{\partial \tau_1'^2} \left[ \psi_+(k, \tau_f) a_+(k, \tau_1') + \psi_-(k, \tau_f) a_-(k, \tau_1') \right] \nonumber\\
&=& s(\tau_0)\, {\cal N}(\tau_0,\tau_2') \tau_0^2 \,
\left[ \psi_+(k, \tau_f) \dot{a}_+(k, \tau_0) + \psi_-(k, \tau_f) \dot{a}_-(k, \tau_0) \right]  \nonumber\\
&+& \int_{\tau_0}^{\tau_f} d\tau_1' \frac{\partial}{\partial \tau_1'}  \left[ s(\tau_1')\, {\cal N}(\tau_1',\tau_2') \tau_1'^2 \right]
\frac{\partial}{\partial \tau_1'} \left[ \psi_+(k, \tau_f) a_+(k, \tau_1') + \psi_-(k, \tau_f) a_-(k, \tau_1') \right] \, .
\end{eqnarray}
The term which is evaluated at $\tau_1' = \tau_f$ vanishes on account of Eq. (\ref{useful_properties_of_a_pm}).  Hence there are no $k$-independent terms that arise from the second time derivative. 

The above analysis shows that the self-correlation of a fluid element may be identified as
\begin{eqnarray}
\langle \delta n(\xi_1,\tau_f) \, \delta n(\xi_2,\tau_f) \rangle_{\rm self}
&=& \frac{s(\tau_f)}{D_Q} \int_{\tau_0}^{\tau_f} d\tau_2' s(\tau_2') \, {\cal N}(\tau_1',\tau_2') \int \frac{dk}{2\pi} \e^{ik(\xi_1-\xi_2)} \nonumber\\
& \times & \left[ \psi_+(-k, \tau_f) a_+(-k, \tau_2') + \psi_-(-k, \tau_f) a_-(-k, \tau_2') \right] \, .
\end{eqnarray}
Using the fact that $s(\tau_2') = s(\tau_f)\tau_f/\tau_2'$, the explicit expression for ${\cal N}$ from Eq. (\ref{calN}), and recognizing that the term in the square bracket above is just $\tilde{G}(k; \tau_f, \tau_2')/ik$, we have
\begin{eqnarray}
\langle \delta n(\xi_1,\tau_f) \, \delta n(\xi_2,\tau_f) \rangle_{\rm self}
&=& \frac{\chi_Q(\tau_f) T_f}{A \tau_f} \, \frac{\tau_f}{\tau_Q} 
\int_{\tau_0}^{\tau_f} \frac{d\tau_2'}{\tau_2'} \left[ {\rm e}^{-(\tau_f - \tau_2')/\tau_Q} - {\rm e}^{-(\tau_f + \tau_2' - 2\tau_0)/\tau_Q} \right]  \nonumber \\
& \times & \int \frac{dk}{2\pi} \e^{ik(\xi_1-\xi_2)} \frac{\tilde{G}(k; \tau_f, \tau_2')}{ik} \, .
\label{selfie}
\end{eqnarray} 
This is the term that ought to be subtracted from the full correlator $\langle \delta n(\xi_1,\tau_f) \, \delta n(\xi_2,\tau_f) \rangle$ to eliminate the self-correlations.  In general it cannot be simplified any further due to the complicated nature of the Green function.  However, it can be calculated in several limits.

When $\tau_Q \rightarrow 0$ the only contribution comes from the upper limit of the integration over $\tau_2'$.  Recall that $\tilde{G}(k; \tau_f, \tau_f) = ik$.  Writing $\tau_2' = \tau_f - \epsilon$, and letting the upper limit of $\epsilon$ go to infinity (because $(\tau_f - \tau_0)/\tau_Q \rightarrow \infty$), we find
\be
\langle \delta n(\xi_1,\tau_f)  \, \delta n(\xi_2,\tau_f) \rangle_{\rm self} = \frac{ \chi_Q(\tau_f) T_f}{A \tau_f}\, \delta (\xi_1 - \xi_2)
\ee
which is exactly the pure diffusion result found above.

For small but nonzero values of $\tau_Q$ one can use the Green function from Eq. (\ref{diffusionG}).  Then a simple calculation gives
\be
\langle \delta n(\xi_1,\tau_f)  \, \delta n(\xi_2,\tau_f) \rangle_{\rm self} = \frac{ \chi_Q(\tau_f) T_f}{A \tau_f}\, 
\frac{v_Q \tau_f}{2D_Q} \exp{\left(-\frac{v_Q \tau_f}{D_Q} |\xi_1 - \xi_2| \right)} \,.
\label{self_smalltau}
\ee
This illustrates the smearing of the Dirac $\delta$-function.  Taking the limit $v_Q \rightarrow \infty$, equivalently $\tau_Q \rightarrow 0$, one recovers the pure diffusion result. 

\section{Numerical Results for Density-Density Correlation Functions}

Before discussing physical observables such as the charge balance functions, let us develop some intuition for how colored noise affects the evolution and development of the collision system.  This is most conveniently done by studying the evolution of the density-density correlation functions $\left< \delta n \delta n \right>$.  In this section we present some numerical results for these density-density correlation functions, showing how they evolve from small $v_Q^2$ to large $v_Q^2$ and eventually to the normal diffusion limit.  These results will be used in the next section to compute the balance functions for various hadrons.  For the sake of definiteness we choose $\tau_0 = 0.5$ fm/c, $T_0 = 350$ MeV, and $T_f = 150$ MeV, 
which imply that $\tau_f = 6.352$ fm/c assuming that the entropy density $s \propto T^3$.  We  also use $D = 0.162$ fm, which is an average over the temperature interval from 150 to 350 MeV taken from Ref. \cite{Aarts2015}.  Unless otherwise noted, we consider only correlation functions for which the self-correlations have been subtracted out by the procedure discussed in the preceding section.

\begin{figure}[h]
\centering
\includegraphics[width=0.6\linewidth]{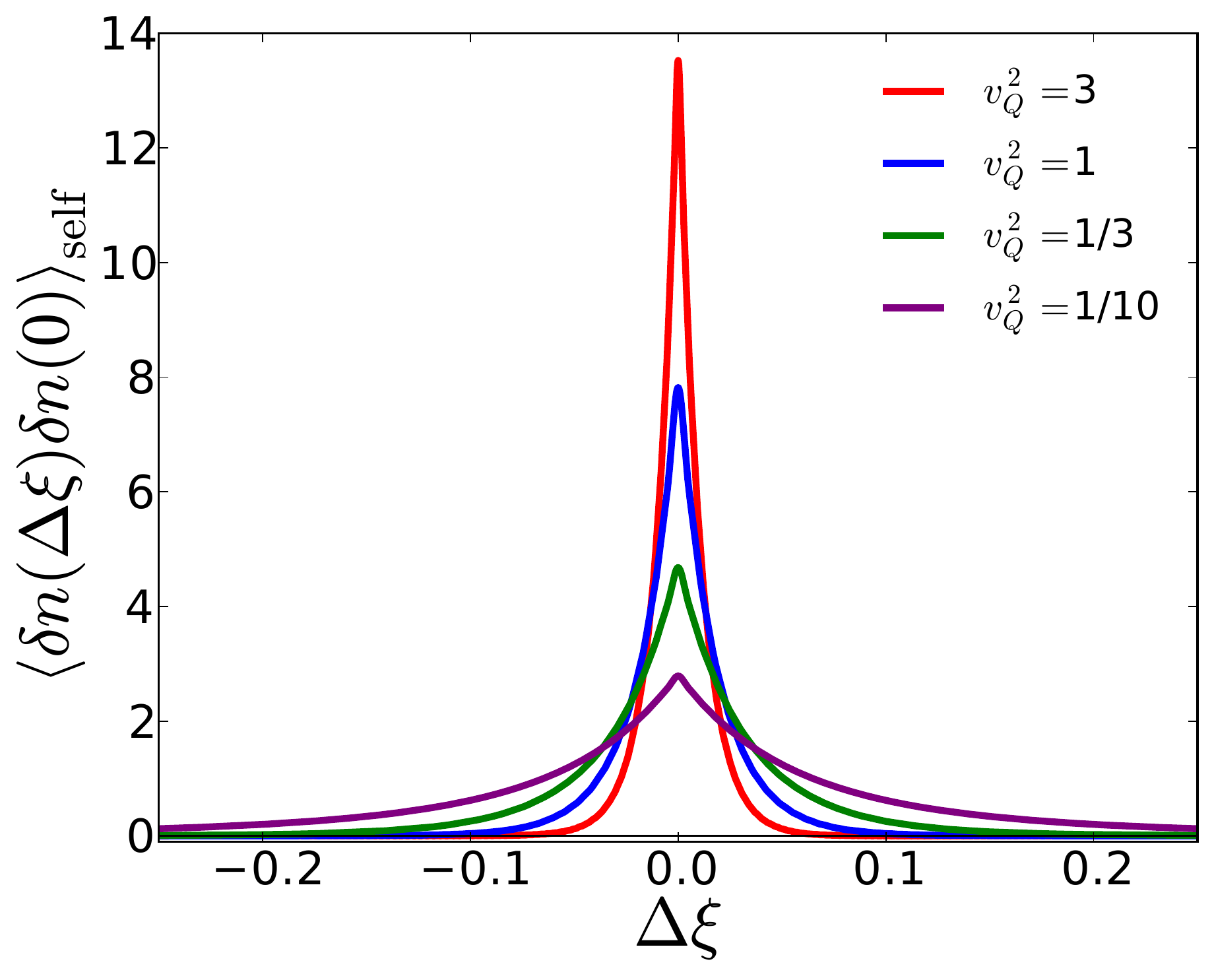}
\caption{(Color online) The density-density self-correlation function versus $\Delta \xi$ for various values of $v_Q^2$ evaluated at the final time $\tau_f$.  For large $v_Q^2$ it approaches the exponential form of Eq. (\ref{self_smalltau}) and eventually a Dirac $\delta$-function.}
\label{self_vs_xi}
\end{figure}

We first consider how the self-correlations change with $v_Q^2$ (or, equivalently, $\tau_Q$).  Figure \ref{self_vs_xi} shows the self-correlation at the final time $\tau_f$ for several values of $v_Q^2$ according to Eq. (\ref{selfie}).  For large values it is well represented by the exponential form of Eq. (\ref{self_smalltau}) with height proportional to $v_Q$ and width inversely proportional to it.  As $v_Q$ becomes smaller the height decreases and the width increases.  In essence the self-correlation is reduced due to the memory effect of having a finite value of $\tau_Q$.

\begin{figure}[ht]
\centering
\includegraphics[height=0.85\textheight,keepaspectratio]{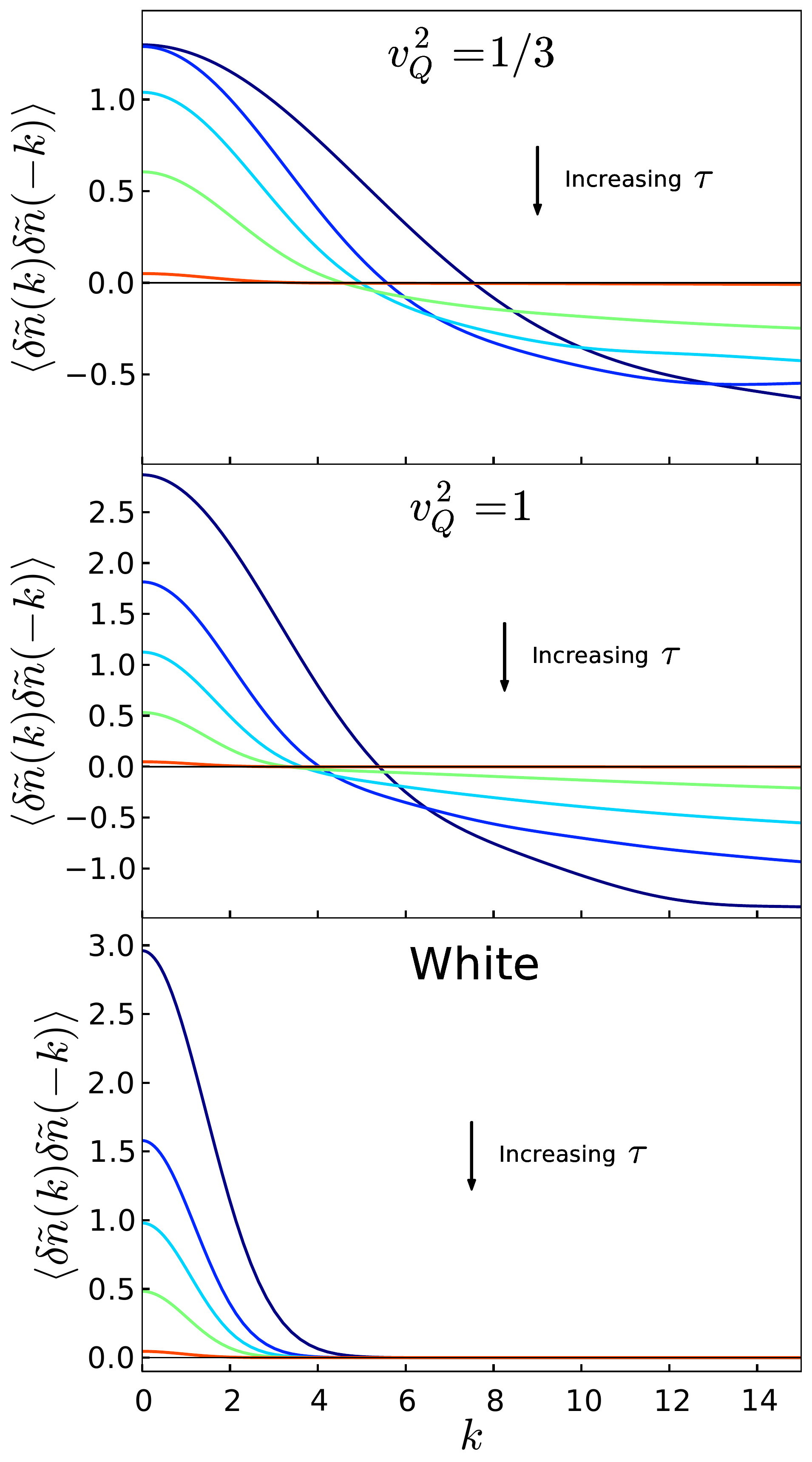}
\caption{(Color online) Time evolution of the density-density correlation function (with self-correlations subtracted) versus $k$.  Top panel: $v_Q^2=1/3$.  Middle panel: $v_Q^2=1$.  Bottom panel: ordinary diffusion with $v_Q^2 \rightarrow \infty$.  The different curves in each panel correspond to elapsed times of 5, 10, 15, 25 and 100\% of the system's lifetime $\tau_f - \tau_0$ = 5.852 fm/c, starting at the top and working down at $k=0$.}
\label{denden_k}
\end{figure}

The panels in Fig. \ref{denden_k} show the time evolution of the density-density correlation function in $k-$space for illustrative values of $v_Q^2$.  The first point to note is that after building up very quickly (we are assuming throughout that there are no initial state correlations or fluctuations) they decrease in time due to the expansion and cooling of the system.  The second point to notice is that for large values of $v_Q^2$ the correlations are essentially Gaussian.  This can easily be seen in the limit of white noise for ordinary diffusion.  For typical values of $k$, say $|k| < 5 - 10$, the correlations are also Gaussian, but for larger values the correlation becomes negative. 

\begin{figure}[bh]
\centering
\includegraphics[width=0.59\linewidth]{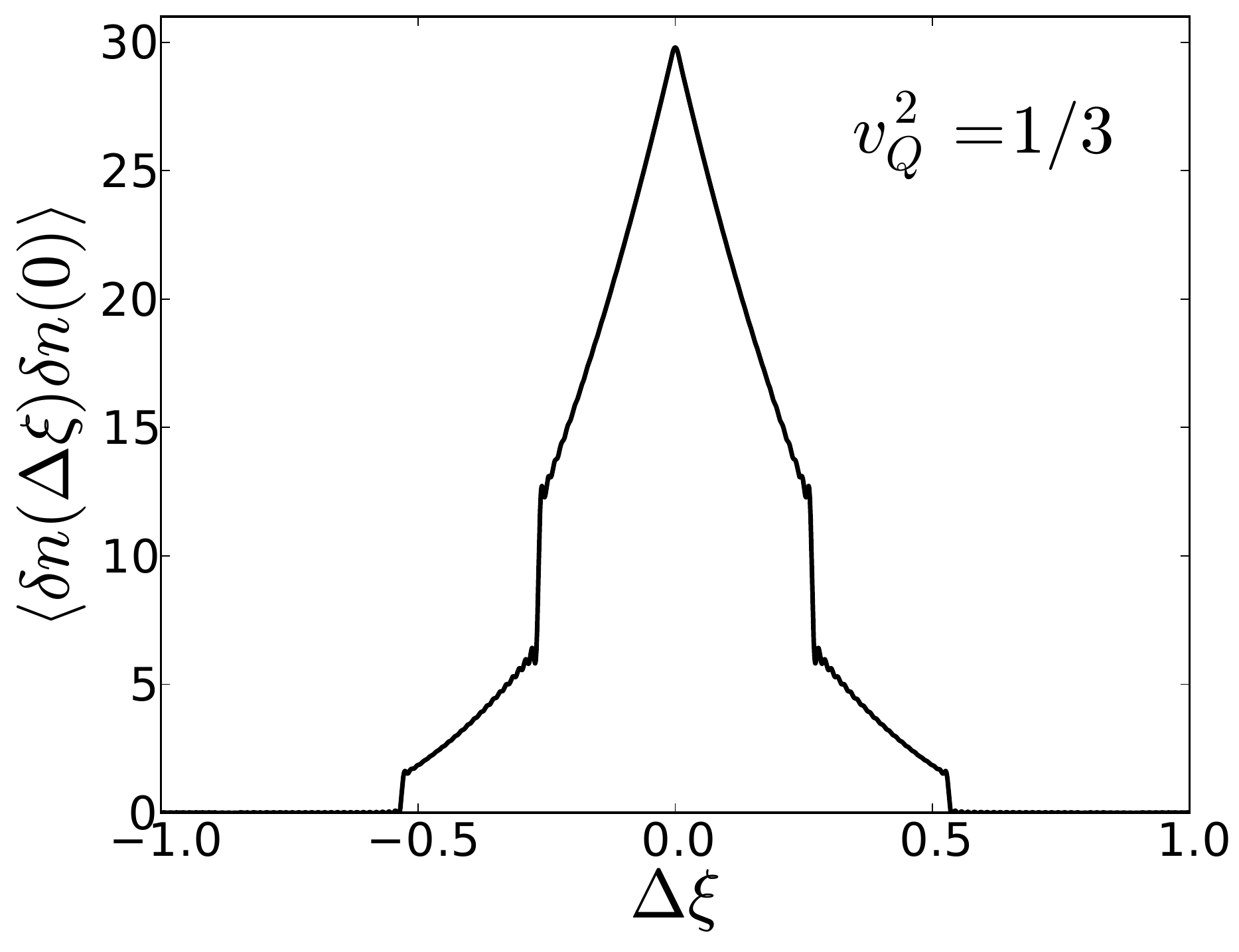}
\includegraphics[width=0.59\linewidth]{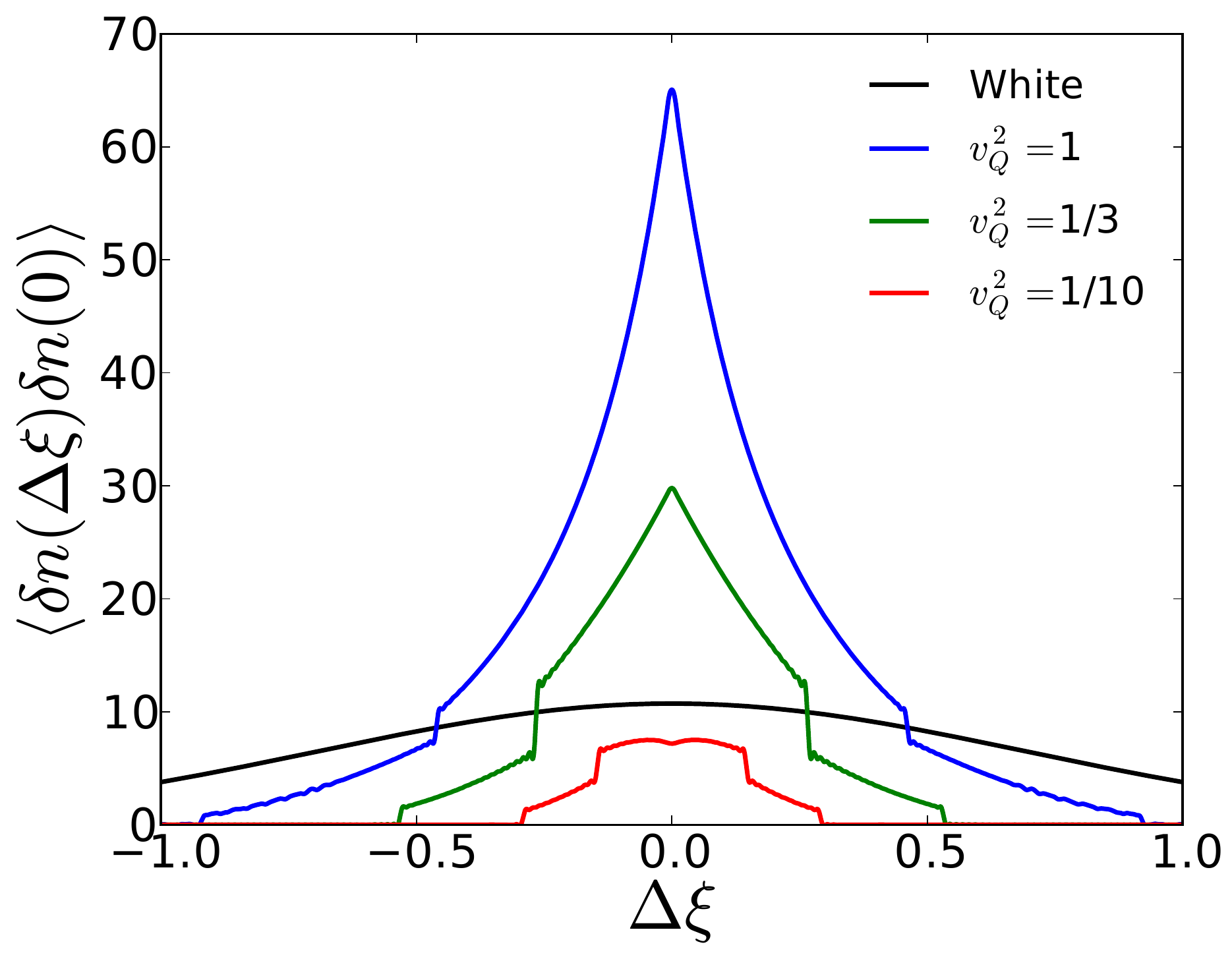}
\caption{(Color online) The regular part of the density-density correlator in $\xi$ space after 5\% of the total expansion time of $\tau_f - \tau_0$ = 5.852 fm/c has elapsed.  The smooth broad curve corresponds to ordinary diffusion with white noise.}
\label{WF_propagation}
\end{figure}

By performing an inverse Fourier transform from $k$ to $\Delta \xi$ we can study how these same correlation functions evolve with time in coordinate space.  We depict this in Fig. \ref{WF_propagation} at a proper time of $\tau =$ 0.793 fm (corresponding to 5\% of the system's total lifetime $\tau_f - \tau_0$ = 5.852 fm/c); in the top panel for $v_Q^2 = 1/3$, and in the bottom panel for several values of $v_Q^2$.  One observes two sets of sharp discontinuities in the dependence of $\langle \delta n (\Delta \xi) \delta n(0) \rangle$ on $\Delta \xi$, both of which reflect the propagation of disturbances through the system with a finite speed.  For fixed $\tau$ and $v_Q$, the discontinuities occur at fixed intervals of $|\Delta \xi| = \xi_s$ and $|\Delta \xi| = 2 \xi_s$, where $\xi_s \equiv v_Q \ln(\tau/\tau_0)$ represents the total distance in space-time rapidity that a disturbance propagates in a time $\tau-\tau_0$.  The bottom panel clearly shows that disturbances in systems with larger $v_Q$ can travel farther than disturbances in systems with smaller $v_Q$.  Note that only the regular part of the correlation function is plotted.  As noted in Ref. \cite{Kapusta2012} there are also singular contributions at the horizons; these contributions are removed by subtracting the large $k$ behavior before doing the Fourier transform.  Of course they will be included when calculating the balance functions in the next section.  The large $k$ behavior for the Green function is calculated anaytically in the appendix.

\begin{figure}[t]
\centering
\includegraphics[width=0.6\linewidth]{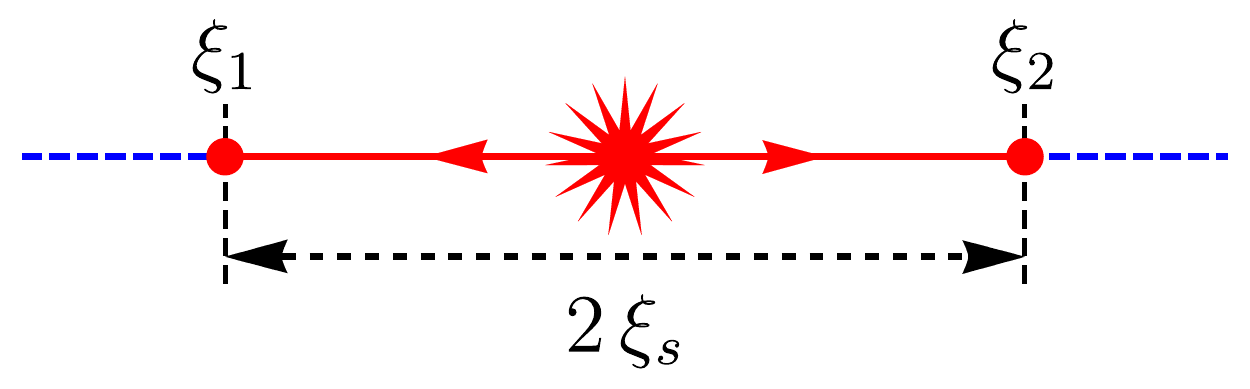}
\includegraphics[width=0.6\linewidth]{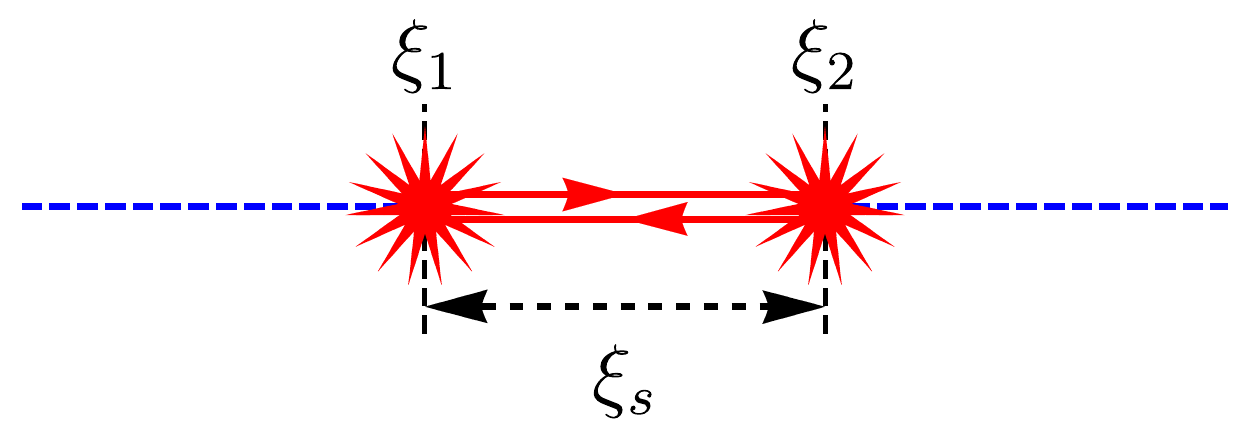}
\caption{(Color online) Schematic depiction of fluctuations and their horizons.}
\label{SchematicCorrelationSources}
\end{figure}

The reason that there are two sets of discontinuities is illustrated schematically in Fig. \ref{SchematicCorrelationSources}.  The upper part shows a fluctuation which had occurred at the midpoint between $\xi_1$ and $\xi_2$.  Those two points will be correlated if their separation is no more than $2 \xi_s$.  The lower part shows a fluctuation which orginated at $\xi_1$ and traveled a distance $\xi_s$, just reaching the point $\xi_2$.  The reverse can also happen.  If a fluctuation occurs at a point to the left of $\xi_1$ or to the right of $\xi_2$ it could not affect both points at the time $\tau$.  

The formalism we presented in the preceding sections therefore incorporates a finite speed of propagation into the standard theory of hydrodynamical fluctuations, and this is clearly born out by a careful analysis of the density-density correlation functions and their time evolution in our simplified model of heavy-ion collisions.  In the next section, we will relate these correlation functions to the charge balance functions and show that the effects of colored noise have important consequences for these physical observables.

\section{Balance Functions}

In this section we calculate the charge balance functions which have been studied extensively elsewhere \cite{Todd2014, Pratt2015}.  We study these quantities in the context of heavy-ion collisions at top RHIC and LHC energies, so that we are justified in taking $\mu_Q = 0$ on average.  Since we focus exclusively on fluctuations of the number density $\delta n$, we need to study how these fluctuations are mapped by the Cooper-Frye formula \cite{Cooper1974} onto the final-state fluctuations which are quantified by the charge balance functions.  This procedure has already been done for our hydrodynamical model in other studies \cite{Kapusta2012, Juan2012, Todd2014}, so we only quote the most important results here.

The distribution of particles along the freeze-out surface $\Sigma_f$ is given by Ref. \cite{Cooper1974}
\be
E \frac{dN}{d^3p} = d \int_{\Sigma_f} \frac{d^3\sigma_{\mu}}{(2\pi)^3} \ p^{\mu} f(\mb{x},\mb{p}) \ ,
\ee
where $d$ is the degeneracy of the particle species under consideration.  We take the distribution function
\be 
f(\mb{x},\mb{p}) = {\rm e}^{-(u \cdot p-\mu)/T} 
\ee
to be the Boltzmann distribution function, where $\mu$ is the chemical potential for that particle, the four-velocity of the fluid cell is
\be 
u^{\mu} =\left(\cosh\xi,0,0,\sinh\xi\right)  \ ,
\ee
and the energy flux through an infinitesimal freeze-out fluid cell is given by
\be 
d^3\sigma_\mu \ p^{\mu} = \tau_f \ d\xi \ d^2x_{\perp} m_\perp \cosh(y-\xi) \ . 
\ee
The variable $y$ represents the particle rapidity
\be 
p^\mu = (m_\perp \cosh y, \mb{p}_\perp, m_\perp \sinh y) \ , 
\ee
where
\be 
m_\perp = \sqrt{m^2+p_\perp^2} 
\ee
is the transverse mass.  The number of particles per unit rapidity is then
\be 
\label{eq:distro} \frac{dN}{dy} =\frac{d A \tau_f }{(2\pi)^3} \ \int d\xi \cosh(y-\xi) \int d^2p_\perp m_\perp 
\exp\left\{-\left[m_\perp\cosh(y-\xi) -\mu \right]/T_f\right\} \ , 
\ee
where the integration over $\mb{x}_{\perp}$ gives the transverse area of the collision $A$.
If we neglect fluctuations by setting $\delta \mu=0$, we get the average of $dN/dy$ as
\be 
\Big\langle \frac{dN}{dy} \Big\rangle = \frac{d A \tau_f}{(2 \pi)^2} \int_{-\infty}^\infty d\xi \ \cosh(y-\xi) \int dp_\perp p_\perp m_\perp 
\exp\left\{-m_{\perp}\cosh(y-\xi)/T_f\right\}  \ .
\ee
In order to perform the integration over $p_\perp$ we use the following formula:
\ba
\label{eq:integration1} \int dp_\perp p_\perp m_{\perp} {\rm e}^{-c m_{\perp}} &= &  \frac{1}{c^3} \ {\rm e}^{-cm} [2+2cm+(cm)^2] \equiv \frac{1}{c^3} \Gamma(3,cm) \ .
\ea
At the freeze-out time we obtain
\be 
\frac{dN}{dy} =\frac{d A \tau_f T_f^3}{4 \pi^2} \int_{-\infty}^{\infty} \frac{dx}{ \cosh^2 x} 
\Gamma \left(3, \frac{m}{T_f} \cosh x \right) \ .
\ee

Now we consider fluctuations of $dN/dy$ and eventually its two-point correlation. To do so, we expand the exponential term
in (\ref{eq:distro}) to first order in fluctuations of $\delta \mu$ around the freeze out value of $\mu_f = 0$:
\be 
\mu = \delta \mu (\tau_f,\xi) \ .
\ee
The Boltzmann factor becomes
\bd
\exp\left\{-[(\cosh(y-\xi) m_\perp-\mu)/T_f]\right\} \rightarrow  \exp\left\{-[m_\perp \cosh(y-\xi)/T_f]\right\}
\left\{ 1 + \frac{\delta \mu (\xi)}{T_f} \right\} \ ,
\ed
where the fluctuations are understood to be evaluated at $\tau_f$.  The fluctuation in the number of particles per unit rapidity is then
\bd
\delta \left( \frac{dN}{dy} \right) = \frac{d A\tau_f }{(2\pi)^3} \int d\xi \cosh(y-\xi) \int d^2 p_\perp m_\perp 
\exp\left\{-m_\perp\cosh(y-\xi)/T_f\right\}\left\{\frac{\delta \mu (\xi)}{T_f} \right\}\ .
\label{eq:fluc} 
\ed
To express this in terms of $\delta n$ we use the fact that
\be
\delta \mu = \frac{\delta n}{\chi_Q} \  \ , 
\ee
where $\chi_{Q} = \left. \partial^2 P(T,\mu)/\partial \mu^2 \right| _{\mu=0}$ is the charge susceptibility discussed above.  By rewriting the equation of state used in Ref. \cite{Juan2012} in terms of the electric charge chemical potential $\mu_Q$, we can write $\chi_{Q}$ explicitly as $\chi_{Q} = \twoth T^2$ when including up, down and strange quarks.

We now perform the integration over $\mb{p}_{\perp}$ with the help of Eq. (\ref{eq:integration1}). The fluctuation of $dN/dy$ reads
\be \delta \left( \frac{dN}{dy} \right) = \frac{dA\tau_f T_f^2}{4\pi^2}  \
 \int d\xi \ \delta n \ F_n (y-\xi) \ .
\ee
Here we have introduced the function
\begin{eqnarray}
F_n (x) \equiv \frac{1}{\chi_{Q} \cosh^2x} \Gamma \left(3,\frac{m}{T_f} \cosh x \right) \ . 
\end{eqnarray}
Finally, we construct the rapidity correlator:
\bd 
\Big\langle \delta \left( \frac{dN}{dy_1} \right) \delta \left( \frac{dN}{dy_2} \right) \Big\rangle
= \left( \frac{dA\tau_f T_f^2}{4\pi^2} \right)^2
\int d\xi_1 \int d\xi_2  F_n(y_1-\xi_1) F_n(y_2-\xi_2) C_{nn}(\xi_1-\xi_2;\tau_f) \ ,
\ed
where
\be 
C_{nn} (\xi_1-\xi_2; \tau_f) = \langle \delta n(\xi_1;\tau_f) \delta n(\xi_2;\tau_f) \rangle_{\rm self}
- \langle \delta n(\xi_1;\tau_f) \delta n(\xi_2;\tau_f) \rangle \ . 
\ee
Note that the self-correlation has been subtracted in the formula for $C_{nn}$.

The appropriate expression for the charge balance function is
\ba
B(\Delta y) & \equiv & \Big\langle \delta \left( \frac{dN}{dy_1} \right) \delta \left( \frac{dN}{dy_2} \right) \Big\rangle \Big\langle \frac{dN}{dy} \Big\rangle^{-1} \ .  \\
&=& \frac{d A \tau_f T_f}{4\pi^2} \frac{C(\Delta y)}{Q \left(m/T_f \right)} \ . 
\ea
Here 
\ba
C(\Delta y) &=& \frac{1}{\tau_f^2}\int dk \ {\rm e}^{ik\Delta y} \tilde{F}_n(k) \tilde{F}_n(-k)
\int_{\tau_0}^{\tau_f} d\tau_1' s(\tau_1')
\int_{\tau_0}^{\tau_f} d\tau_2' s(\tau_2') \, {\cal N}(\tau_1',\tau_2') \nonumber \\
&\times& \left[ \tilde{G}(k; \tau_f,\tau_1') \tilde{G}(-k; \tau_f,\tau_2')\Big|_{\rm self} - \tilde{G}(k; \tau_f,\tau_1') \tilde{G}(-k; \tau_f,\tau_2') \right]
\ea
and
\be 
Q \left(m/T_f \right) \equiv \int_{-\infty}^{\infty} \frac{dx}{\cosh^2 x} \Gamma \left( 3,\frac{m}{T_f} \cosh x\right) \ . 
\ee

\begin{figure}
\centering
\includegraphics[width=0.8\linewidth]{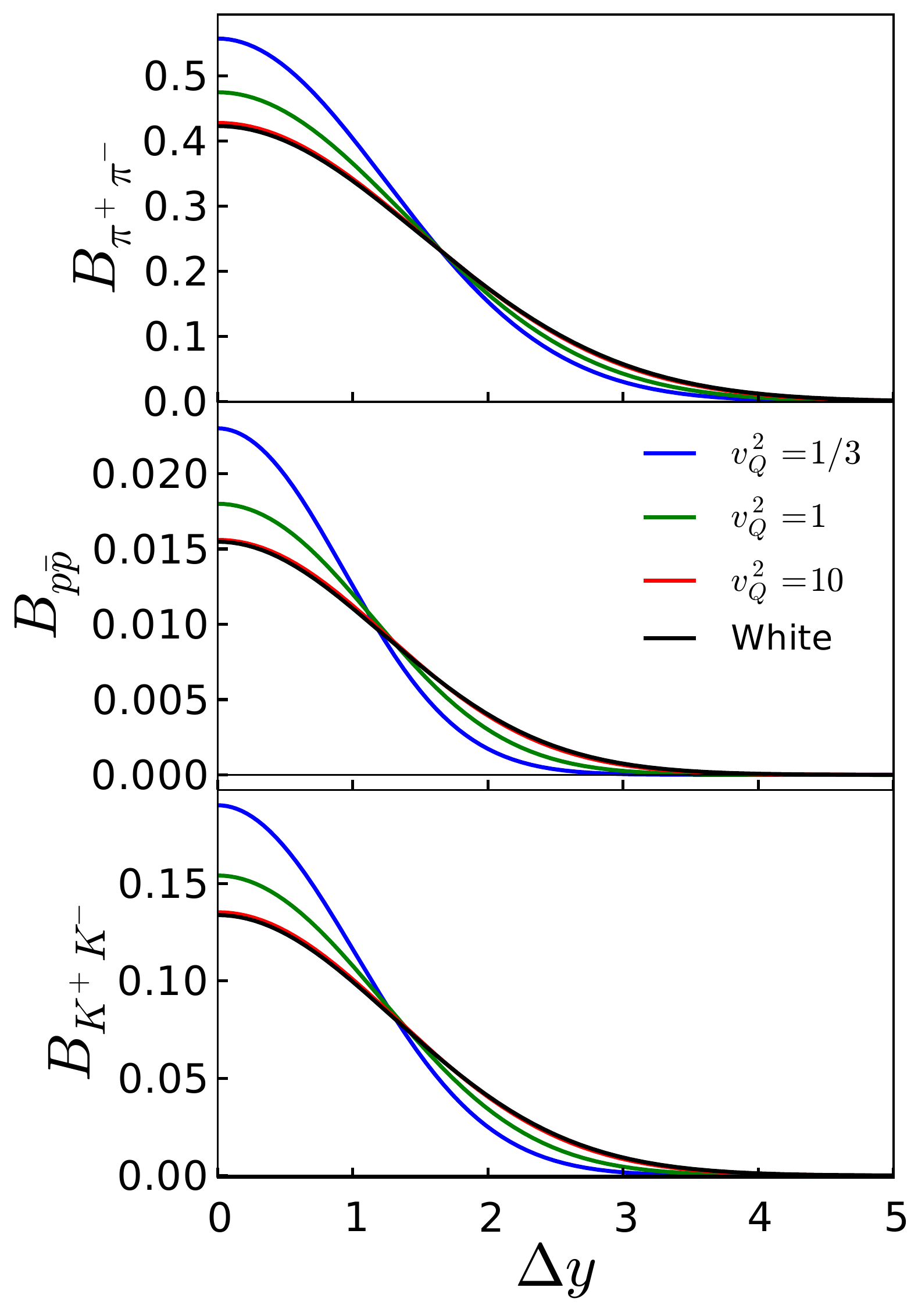}
\caption{(Color online) Balance functions for pions, protons, and kaons.  Looking at $\Delta y = 0$ the curves correspond to $v_Q^2$ = 1/3 (top), 1 (middle) and 10 (bottom), with 10 being indistinquishable from the case of white noise in the ordinary diffusion equation.}
\label{BFresults}
\end{figure}

The balance functions are shown for pions, protons, and kaons in Fig. \ref{BFresults} for various choices of $v_Q^2$.  We see that the balance functions are systematically enhanced at $\Delta y = 0$ and are narrower for smaller versus larger values of $v_Q^2$.  These are natural consequences of the increasing efficiency with which fluctuations are propagated through the system as $v_Q^2$ is increased.  Moreover, as $v_Q^2$ is increased we find that the correlations tend to the case of white noise, as to be expected.   

\newpage

\section{Comparison of Gurtin-Pipkin Equation with AdS/CFT: Estimation of Parameters}

The anti-de Sitter space/conformal field theory (AdS/CFT) correspondence is often used as a guide to the values of transport coefficients in the strongly coupled sector of QCD.  The best known of these is the suggestion that the ratio of the shear viscosity to entropy density $\eta/s$ has a universal lower bound of $1/4\pi$ \cite{Son2005}.  A good overview is provided in Ref. \cite{adscftBook}.  In this section we make a tentative estimation of the parameters appearing in relativistic causal diffusion.

Reference \cite{Policastro2002} studied the correlator of R-charge currents in ${\cal N} = 4$ Super Yang-Mills theory.  They found a pole in the current-current correlation function corresponding to a pure diffusion mode under the assumption of small frequency $\omega$ and wave-number $k$.  It is
\be
\omega = -iDk^2 + \cdot\cdot\cdot
\ee
with $D = 1/2\pi T$.  Reference \cite{Nunez2003} studied the analytic structure of the correlator when $k=0$ but the magnitude of $\omega$ is arbitrary.  They found additionally a pair of complex poles located at 
\be
\omega(k=0) = \left(\pm n - in\right) 2\pi T \,, \;\;\;\; n = 1, 2, 3, ...
\label{adscftpoles}
\ee
Thus the correlator has one pure diffusion mode and a tower of a pair of complex modes.  This Kaluza-Klein tower is characterstic of AdS/CFT excitations.  

Matching all of these with a conserved current cannot be done using a differential equation with a finite number of derivatives.  Here we only try to match the lowest energy modes.  This implies an equation involving third order derivatives to give three poles: one pure diffusive and a pair of complex poles.  The Cattaneo equation has only second derivatives, so we go to the Gurtin-Pipkin equation which has third derivatives \cite{GP,Clint2014}.

Going to third order in derivatives results in an equation first applied to the problem of heat conduction by Gurtin and Pipkin \cite{GP}.  It is
\be
\left[ \frac{\partial}{\partial t} - D\nabla^2 + \tau_1 \frac{\partial^2}{\partial t^2} + \tau_2^2 \frac{\partial^3}{\partial t^3}
- \tau_3' D\frac{\partial}{\partial t}\nabla^2  \right] n = 0 \,.
\ee
This equation is hyperbolic. The cubic equation following from this in frequency and wave-number is
\be
\tau_2^2 \omega^3 + i \tau_1 \omega^2 - (1 + \tau_3' D k^2) \omega - i D k^2 = 0 \,.
\label{cubic}
\ee
High frequency waves travel with speed $v = \sqrt{\tau_3' D/\tau_2^2}$.   It follows from the current
\be
J^{\mu} = n u^{\mu} + \sigma T \Delta^{\mu} \frac{1 + \tau_4(u \cdot \partial)}{1+\tau_1(u \cdot \partial) + \tau_2^2(u \cdot \partial)^2
 + \tau_3 D \Delta^2} \left(\frac{\mu}{T}\right)
\ee
where the differential operator in the denominator is to be understood as its Taylor series expansion.  Note that there are four time constants in the current as $\tau_3' = \tau_3 + \tau_4$.  Obviously, setting $\tau_2 = \tau_3 = \tau_4 = 0$ results in the Cattaneo equation, and further setting $\tau_1 = 0$ results in the ordinary diffusion equation.  

When $k=0$ the pair of complex poles from Eq. (\ref{cubic}) are
\be
\omega_{\pm} = \pm \frac{1}{\tau_2} \sqrt{1 - \left(\frac{\tau_1}{2\tau_2}\right)^2} - i \frac{\tau_1}{2\tau_2^2}
\ee
In order to reproduce \ref{adscftpoles} with $n=1$ requires that $\tau_1 = 1/2\pi T$ and $\tau_2 = \tau_1/\sqrt{2}$.  Reference \cite{Nunez2003} calculated the dispersion relation for these poles numerically.  At large $k$ the real parts are $\omega = k + \cdot\cdot\cdot$ so that $v = \sqrt{\tau_3' D/\tau_2^2} = 1$.  Hence we infer that $\tau_3' = 1/4\pi T$.

It is not our goal here to make a detailed comparison of the poles and residues arising in AdS/CFT and the Gurtin-Pipkin equation.  We just note that the correction to the diffusive mode at the next order in $k$ from AdS/CFT is \cite{Policastro2002}
\be
\omega = -iDk^2 \left( 1 + \frac{\ln 2 \, k^2}{(2\pi T)^2} \right)+ \cdot\cdot\cdot \,.
\ee 
We can determine the order $k^4$ term from Eq. (\ref{cubic}) to be
\be
\omega = -iDk^2 \left( 1 + (\tau_1 - \tau_3') D k^2 \right) = -iDk^2 \left( 1 + \frac{(1 - \thalf) k^2}{(2\pi T)^2} \right) \,.
\ee
This gives a coefficient of 1/2 versus $\ln 2$ in the $k^4$ term which is close but not identical.  However, it is amusing to note the expansion $\ln 2 = 1 - \thalf + \oneth - \cdot\cdot\cdot$.  It may be that the AdS/CFT result includes contributions from the Kaluza-Klein tower of excitations and that the Gurtin-Pipkin equation only captures the first two terms.  But that is only speculation.

The Cattaneo current does not keep the higher order derivatives appearing in the Gurtin-Pipkin current.  The inference is that $\tau_Q = \tau_1 = 1/2\pi T$ and $v_Q^2 = D_Q/\tau_Q = 1$ which are effectively consistent with the AdS/CFT results.

\section{Conclusions}
\label{sec:conclusion}

In this paper we used a form of the electric current in matter that propagates signals at finite speed and includes a noise term following from the fluctuation-dissipation theorem.  This may be referred to as the Cattaneo current as it follows from his approach to heat conduction.  As is well known, the usual diffusion equation propagates signals instantaneously, and this causes problems when modeling high energy heavy ion collisions.  Our goal was to understand the underlying physics in a simple well defined problem, namely, the 1+1 dimensional Bjorken hydrodynamical model.  Any more realistic numerical modeling of heavy ion collisions must be able to reproduce the semi-analytical results obtained here.  Apart from the diffusion constant $D_Q$ there also appears a characteristic time scale $\tau_Q$.  Propagation of signals less than the speed of light requires that $\tau_Q > D_Q$ as $v_Q^2 = D_Q/\tau_Q$, whereas ordinary diffusion corresponds to $\tau_Q \rightarrow 0$.  Our numerical study assumed that both $D_Q$ and $\tau_Q$ were temperature independent; however, that assumption may be relaxed at the expense of somewhat more involved solutions of the differential equation (\ref{gencon}).  The solutions would still involve Kummer and Whittaker functions but with more complicated arguments.

We then used these results to compute the balance functions for electrically charged hadrons in the central rapidity region in very high energy nuclear collisions.  As one would expect intuitively, limiting the speed of propagation of fluctuations leads to a narrowing of the balance functions and a corresponding increase in their height at small rapidity separation.  However, the magnitude of these efects are somewhat reduced by the thermal smearing of hadrons emitted from the fluid elements. 

As part of the analysis we subtracted the self-correlations from the same or nearby fluid elements in order to be consistent with how experimental measurements are done.  As long as $\tau_Q$ is very small compared to the lifetime of the system $\tau_f-\tau_0$ this is a fairly well defined procedure.  If it is not, then we are challenged to separate out self-correlations and how this relates to observable quantities.

Clearly there is much work to be done.  One step to include the analogous baryon current into 3+1 dimensional fluid models of heavy ion collisions has been reported in Ref. \cite{Chun}.  Incorporation of noise was not included there and remains a notable challenge.  Fortunately data is available from experiments at both RHIC \cite{STAR} and LHC \cite{ALICE}.

\newpage

\acknowledgments

We thank S. Pratt, T. Springer, A. Kamenev, S. Gavin, Y. Yin, M. Martinez, D. Bazow, and J. Vinals for discussions.  This work was supported by the U.S. DOE Grant No. DE-FG02-87ER40328.

\appendix

\section{Large k Limit}

To extract the singularities we shall perform a Laurent expansion of the homogenous solutions for large values of $K \equiv v_Q k$.  We assume that asymptotically the solution can be written as
\be
\psi_+ = x^{-1/2} \, {\rm e}^{-x/2} \, {\rm e}^{iK\ln x} \left[ 1 + \sum_{n=1}^{\infty} \frac{i^n g_n(x)}{K^n} \right]
\ee
and substitute this into the homogeneous  equation \ref{homoeq}.  (The factor of $i^n$ is inserted because it will turn out that the $g_n$ are real.)  To order 1 we find that
\ba
\dot{g}_1 &=& -\frac{x}{8} - \frac{1}{2} - \frac{1}{8x} \nonumber \\
g_1 &=& -\frac{x^2}{16} - \frac{x}{2} - \frac{1}{8} \ln x + c_1 \label{g1}
\ea
where $c_1$ is a constant.  For $n \ge 1$
\be
\dot{g}_{n+1} = \frac{x}{2} \ddot{g}_n + \frac{1}{2} \dot{g}_n - \left(\frac{x}{8} + \frac{1}{2} + \frac{1}{8x} \right) g_n \,.
\label{recursion}
\ee
These equations can be solved to any desired order.  Note, however, that a constant of integration $c_n$ will appear at each order.  They are associated with the logarithms that follow solely from the expansion of
\bd
x^{\lambda_+} = \exp(iK \sqrt{1-1/4K^2} \ln x) 
\ed
for large $K$.  All $c_n$ should be chosen to be zero to agree with the known asymptotics of the Whittaker function.  For example, the next term is
\be
g_2 =  \frac{x^4}{512} + \frac{x^3}{32} + \frac{x^2}{16} - \frac{x}{4} 
+ \frac{x^2}{128} \ln x + \frac{x}{16} \ln x + \frac{1}{128} \ln^2 x \,.
\ee
Alternatively, one may use the representation 
\be
\psi_+ = x^{-1/2} \, {\rm e}^{-x/2} \, {\rm e}^{i K \sqrt{1-1/4K^2} \ln x} \left[ 1 + \sum_{n=1}^{\infty} \frac{i^n p_n(x)}{K^n} \right] \,.
\ee
This separates the log contributions which come solely from the exponential.  The $p_n$ are determined by
\be 
\sum_{n=1}^{\infty} \frac{i^n}{K^n} \left[ \ddot{p}_n + \frac{1}{x} \left( 1 + 2iK \sqrt{1-1/4K^2} \right) \dot{p}_n
- \left(\frac{1}{x} + \frac{1}{4} \right) p_n \right] = \frac{1}{x} + \frac{1}{4} \,.
\ee
The first few terms are
\ba
p_1 &=& - \frac{x^2}{16} - \frac{x}{2} \nonumber \\
p_2 &=& \frac{x^4}{512} + \frac{x^3}{32} + \frac{x^2}{16} - \frac{x}{4}
\ea
subject to the condition that $p_n(0) = 0$.

Now let us return to the asymptotics of the Green function.  Define
\be
\Phi(x) \equiv 1 + \sum_{n=1}^{\infty} \frac{i^n g_n(x)}{K^n} \,.
\ee
Note that $\psi_+ = \psi$ and $\psi_- = \psi^*$ when $K^2 > 1/4$.  Then the numerator of \ref{Green} is
\ba
&&\psi(x) \dot{\psi}^*(x') -  \psi^*(x) \dot{\psi}(x') = (xx')^{-1/2} {\rm e}^{-(x+x')/2} \nonumber \\
&&\times \left\{ {\rm e}^{iK\ln (x/x')} \Phi(x) 
\left[ \left(- \frac{iK}{x'} - \frac{1}{2x'} - \frac{1}{2} \right) \Phi^*(x') + \dot{\Phi}^*(x') \right] \right. \nonumber \\
&&- \left. {\rm e}^{-iK\ln (x/x')} \Phi^*(x) 
\left[ \left( \frac{iK}{x'} - \frac{1}{2x'} - \frac{1}{2} \right) \Phi(x') + \dot{\Phi}(x') \right] \right\} \,.
\ea
When $x=x'$ this simplifies to
\be
\psi(x') \dot{\psi}^*(x') -  \psi^*(x') \dot{\psi}(x') =
- \frac{2i}{x'} {\rm e}^{-x'} \left[ \frac{K}{x'} \left( \Phi_R^2 + \Phi_I^2 \right)
+ \Phi_R \dot{\Phi}_I - \Phi_I \dot{\Phi}_R \right] \,,
\ee
where $\Phi_R$ and $\Phi_I$ are the real and imaginary parts of $\Phi$ evaluated at $x'$.  This means that $\tilde{G}$ has the form $K {\rm e}^{\pm iK\ln (x/x')}$ times a power series in $1/K^n$ starting with $n=0$.  When computing an equal-time correlator $\tau_1 = \tau_2 = \tau_f$ as in equations \ref{corrk} or \ref{corrxi} there will be delta functions and their first and second derivatives plus step functions.

After some lengthy calculation we find the Green function to be
\ba
\tilde{G}(k;x,x') &=& ik \left( \frac{x'}{x} \right)^{1/2} {\rm e}^{-(x-x')/2} \biggl\{ \cos(v_Q k L) + \frac{\sin(v_QkL)}{v_Qk} B_1(x,x') \nonumber \\
&& - \frac{\cos(v_QkL)}{v_Q^2k^2} B_2(x,x') + {\cal O}(1/k^3) \biggr\}
\ea
with
\ba
B_1(x,x') &=&  g_1(x') - g_1(x) + \thalf x' + \thalf \nonumber \\
B_2(x,x') &=&  \thalf \left[ g_1(x') - g_1(x) \right]\left[ 2 g_1(x') + x' + 1 \right] - \left[ g_2(x') - g_2(x) \right] 
\ea
and we have defined $L \equiv \ln (x/x')$.  Note that $\tilde{G}(k;x,x) = i k$ order by order in the series expansion.

\end{document}